%% file: fapi.tex
\documentclass[%
 aip,
amsmath,amssymb,
 reprint,%
]{revtex4-1}

\usepackage[dvipdfmx]{graphicx}
\usepackage{dcolumn}
\usepackage{multirow}
\usepackage{bm}

\usepackage[utf8]{inputenc}
\usepackage[T1]{fontenc}
\usepackage{mathptmx}

\begin{document}


\title{Photo-excited charge carrier lifetime enhanced by slow cation molecular dynamics in lead iodide perovskite FAPbI$_3$}

\author{M. Hiraishi}
\author{A. Koda}
\author{H. Okabe}
\author{R. Kadono}  \email{ryosuke.kadono@kek.jp}
\affiliation{ 
Muon Science Laboratory, 
Institute of Materials and Structural Science, 
High Energy Accelerator Research Organization (KEK), 
203-1, Shirakata, Tokai, Naka, Ibaraki 319-1106, Japan
}%

\author{K. A. Dagnall}
\author{J. J. Choi}
\affiliation{ 
Department of Chemical Engineering, University of Virginia, Charlottesville, Virginia 22904,~USA
}%

\author{S.-H. Lee}
\affiliation{ 
Department of Physics, University of Virginia, Charlottesville, Virginia 22904, USA
}%

\date{\today}

\begin{abstract}
Using muon spin relaxation ($\mu$SR) measurements on formamidinium lead iodide [FAPbI$_3$, where FA denotes HC(NH$_2)_2$] we show that, among the five structurally distinct phases of FAPbI$_3$ exhibited through two different temperature hysteresis, the reorientation motion of FA molecules is quasi-static below $\approx50$ K over the time scale of 10$^{-6}$ s in the low-temperature (LT) hexagonal (Hex-LT, $<160$ K) phase which has relatively longer photo-excited charge carrier lifetime ($\tau_{\rm c}\sim$10$^{-6}$ s). In contrast, a sharp increase in the FA molecular motion was found above $\approx50$ K in the Hex-LT phase, LT-tetragonal phase (Tet-LT, $<140$ K), the high-temperature (HT) hexagonal phase (Hex-HT, 160--380 K), and the HT-tetragonal phase (Tet-HT, 140--280 K) where $\tau_{\rm c}$ decreases with increasing temperature. More interestingly, the reorientation motion is further promoted in the cubic phase at higher temperatures ($>380/280$ K), while $\tau_{\rm c}$ is recovered to comparable or larger than that of the LT phases. These results indicate that there are two factors that determine $\tau_{\rm c}$, one related to the local reorientation of cationic molecules that is not unencumbered by phonons, and the other to the high symmetry of the bulk crystal structure.
\end{abstract}

\maketitle
\section{Introduction}
Hybrid organic-inorganic perovskites (HOIPs) are promising for application in solar cells \cite{Kojima:09} and various other optical devices due to their low production cost \cite{Park:16,Cai:17} and excellent optoelectronic properties \cite{Kanemitsu:18,Schmidt:21}. The high photoelectric conversion efficiency of more than 25\% shown as a solar cell material \cite{Kim:20} is attributed to the long photo-excited charge carrier lifetime ($\tau_{\rm c}\ge10^{-6}$ s), which translates into a large carrier diffusion length despite the modest carrier mobility \cite{Brenner:16}. Several microscopic mechanisms have been proposed as intrinsic factors for the unusually long $\tau_{\rm c}$, including ferroelectric domain deformation \cite{Frost:14a,Kutes:14,Strelcov:17}, the Rashba effect \cite{Zheng:15,Etienne:16}, photon recycling \cite{Yamada:17}, and large polarons \cite{ChenY:16,Zhu:16}. In particular, since the conversion efficiency of all-inorganic perovskite solar cells was well below that of HOIPs \cite{ChenCY:17,Nam:17,Sutton:16}, the presence of organic cations was initially thought to be crucial to achieving higher efficiency in solar cells. Recently, however, the efficiency of all-inorganic perovskites has also improved significantly to over 20\% \cite{Wang:23}, and there is an increasing need to clarify the microscopic relationship between organic cations and solar cell efficiency in order to obtain guidelines for future material design and selection.

Among the aforementioned microscopic mechanisms, three theoretical models have been proposed that involve the electric polarization of cation molecules. First, it is predicted that nanoscale ferroelectric domains formed by an array of organic cations spatially separate photo-excited electrons and holes, thereby reducing their recombination \cite{Frost:14a,Frost:14b,Kutes:14,Liu:15}. Second, in the Rashba effect model \cite{Etienne:16,Hutter:17,Zheng:15}, the electronic bands of inorganic atoms, which are accompanied by spin-orbit interactions, are split by the electric field from the cation molecule to create an effective indirect band gap, thereby increasing carrier longevity. The third model is that organic cations reorient locally in response to the presence of photo-excited carriers, resulting in the formation of large polarons \cite{ChenY:16,Zhu:16,ZhuXY:15}. In the last case, carriers screened by local polarization of organic cations would be protected from scattering by defects and phonons, leading to extended $\tau_{\rm c}$. In order to examine these models, it is important to experimentally elucidate the microscopic correlations between organic cation motions and $\tau_{\rm c}$.

We recently performed muon spin relaxation ($\mu$SR) measurements on methylammonium lead iodide (MAPbI$_3$, where MA denotes CH$_3$NH$_3$) and found that the $\mu$SR linewidth ($\Delta_{\rm MA}$) determined by the magnetic dipolar fields exerted from the nuclear magnetic moments of the MA molecules varies in proportion to the relaxation time ($\tau_{\rm r}$) of molecular reorientation induced by thermal agitation, indicating that $\Delta_{\rm MA}$ is a good probe of local molecular motion over a time range of $10^{-11}<\tau_{\rm r}<10^{-8}$ s\: \cite{Koda:22}. Moreover, the temperature dependence of  $\Delta_{\rm MA}$ (which turned out to be non-monotonic) is also proportional to the photo-luminescence (PL) lifetime ($\tau_{\rm PL}$, $\propto\tau_{\rm c}$), suggesting that $\Delta_{\rm MA}\propto\tau_{\rm r}\propto\tau_{\rm PL}$\:\cite{Koda:22}. 
Considering that the local electric dipole (${\bm P}$) of the MA molecule seen from carriers is effectively reduced to $\overline{{\bm P}}$ by motional averaging as in the case of $\Delta_{\rm MA}$ (i.e., $\overline{{\bm P}}\propto\Delta_{\rm MA}$), this is interpreted that the longer $\tau_{\rm r}$, the larger $\overline{{\bm P}}$ is to promote the formation of large polarons by allowing local dielectric response of the molecules. 
In other words, it is suggested that for local reorientation of cationic molecules, the disadvantage associated with the disturbance of reorientation motion by phonons (shortening of $\tau_{\rm r}$) at high temperatures is greater than the advantage of increased freedom of reorientation motion due to structural phase transition.  This is in contrast to the intuitive expectation where only the latter effect would be considered, and awaits further experimental verification as it needs to be carefully examined with similar materials.

Here, we report our $\mu$SR study on formamidinium lead iodide [FAPbI$_3$,  where FA denotes HC(NH$_2)_2$] to elucidate the local FA molecular motion vs.~temperature in detail. We found a similar correlation between the $\mu$SR linewidth and $\tau_{\rm PL}$ as observed in the case of MAPbI$_3$. Specifically, among the five structurally distinct phases exhibited by FAPbI$_3$ through two different temperature hysteresis, the reorientation motion of FA molecules is quasi-static below $\approx50$ K in the low-temperature (LT) hexagonal (Hex-LT, $<160$ K) phase which exhibits relatively long $\tau_{\rm PL}$ ($\sim$10$^{-6}$ s). In contrast, a sharp increase in the FA molecular motion was found above $\approx50$ K in the Hex-LT phase, LT-tetragonal phase (Tet-LT, $<140$ K), the high-temperature (HT) hexagonal phase (Tet-HT, 160--380 K), and the HT-tetragonal phase (Tet-HT, 140--280 K) where $\tau_{\rm PL}$ is reduced with increasing temperature. More interestingly, in the cubic phase ($>380/280$ K) where the FA molecular motion is further accelerated by thermal agitations, $\tau_{\rm PL}$ is recovered to as large as or larger than that of the LT phases. These results suggest that $\tau_{\rm c}$ is determined by both the relatively slow reorientation motion of cationic molecules (which is not subject to strong thermal disruption) and unknown factors associated with the cubic crystal structure.

\section{$\mu$SR experiment and DFT calculation}
\begin{figure}[t]
        \begin{center}
                \includegraphics[width=0.8\linewidth,clip]{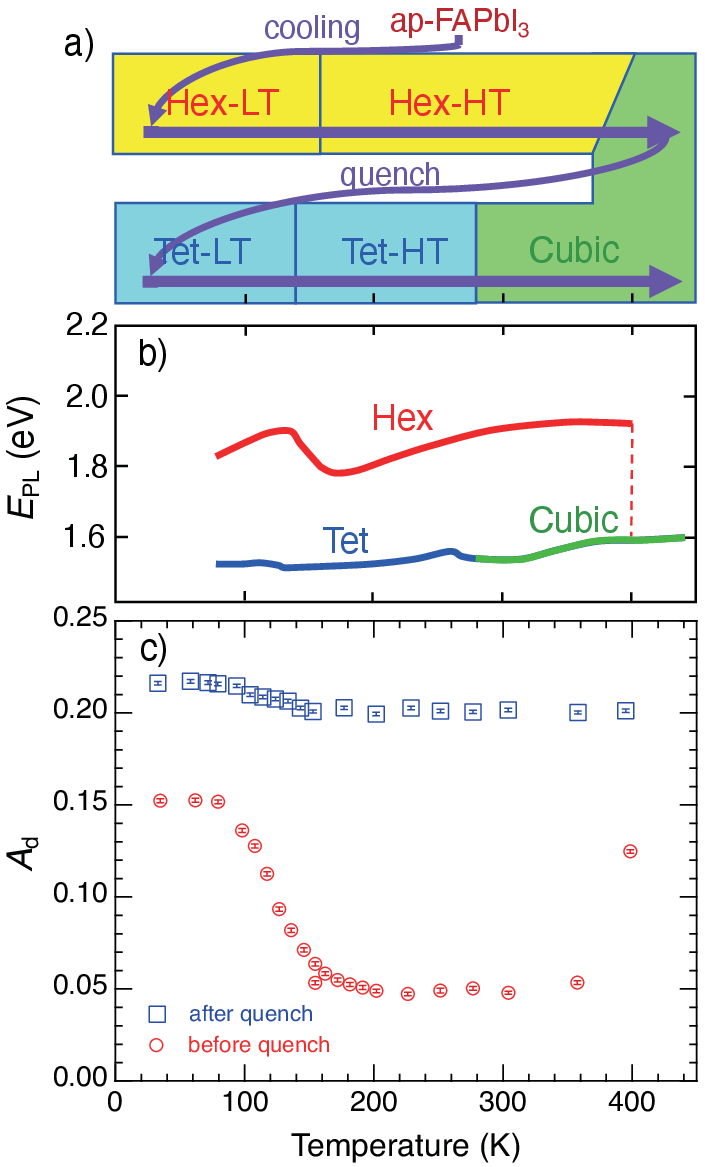}
\caption{(a) The temperature variation sequence for $\mu$SR measurements in FAPbI$_3$, (b) photo-luminescence (PL) energy inferring the band gap (quoted from Ref.~[\onlinecite{ChenT:17}]), and (c) the initial asymmetry of the $\mu$SR spectra under a transverse field of 2.0 mT. Hex-LT/HT denotes low/high-temperature hexagonal phase, and Tet-LT/HT denotes low/high-temperature tetragonal phase. The as-prepared sample (Hex-HT) was cooled down to the base temperature, and $\mu$SR measurements were performed with increasing temperature to reach the cubic phase, then the sample was quenched by rapid cooling for the $\mu$SR measurements in the tetragonal phase.
}
\label{phase}
\end{center}
\end{figure}

The $\mu$SR experiments were performed using the ARTEMIS spectrometer installed in the S1 area at the Materials and Life Science Experimental Facility in J-PARC, where a nearly 100\% spin-polarised pulsed $\mu^+$ beam (25 Hz, with the full width at half-maximum of 80 ns and a momentum of 27 MeV/c) was transported to a powder sample ($\sim$2 g, packed in a disk shape using aluminum foil) mounted on a silver sample holder attached to a He gas-flow cryostat for controlling temperature. $\mu$SR spectra [time-dependent positron asymmetry, $A(t)$] were measured in the two-step temperature change sequence shown in Fig.~\ref{phase}(a). FAPbI$_3$ synthesized at ambient temperature has the most stable hexagonal (Hex-HT) crystal structure, but when the temperature is raised above $\sim$380 K, it transitions to a metastable cubic phase. Then, upon rapid cooling down from this cubic crystal structure, FAPbI$_3$ does not return to the hexagonal crystal structure, but undergoes a structural phase transition to the tetragonal crystal structure. Furthermore, below 160 K and between 160 and 280 K, the FA molecule assumes two different tetragonal structures  (Tet-LT and Tet-HT) associated with the rotational degrees of freedom. Considering this complex phase diagram, $\mu$SR measurements in the first sequence were performed on the as-prepared (ap-) sample after cooling down from $\sim$300 K to the lowest temperature ($\sim$50 K) to follow the path of hexagonal to cubic transition. The sample was then quenched by rapid cooling from 417 K to $\sim$40 K, and $\mu$SR measurements across the tetragonal-to-cubic transition were performed.

The DFT calculations were performed to investigate the local structure of H defects (mimicking Mu) using the projector augmented wave approach \cite{Kresse:99} implemented in the Vienna {\it ab initio} simulation package (VASP) \cite{Kresse:96} with the Perdew-Burke-Ernzerhof (PBE) exchange correlation potential\cite{Perdew:96}, where the lattice parameters reported in the literature were adopted\cite{ChenT:16}. The cutoff energy for the plane-wave basis set was  400 eV. The distribution of thelocal magnetic field at the muon sites was calculated using Dipelec program \cite{Kojima:04}. Crystal structures were visualized using the VESTA program \cite{Momma:11}.

\section{Result}
In general, $\mu^+$ implanted into a material behaves as a light isotope of hydrogen (denoted hereafter by the element symbol Mu). Especially in insulators, it is known that $\mu^+$ can interact with carriers generated by the electronic excitation associated with the implantation and can take on several metastable relaxed-excited states, including neutral hydrogen atom-like state (Mu$^0$). The specific electronic state including its valence is determined by the impurity level associated with Mu and the band structure of the host material\cite{Hiraishi:22}. Therefore, in order to use Mu as a probe, it is often crucial to know the electronic state of Mu in the target material. To this end, $\mu$SR measurements under a weak transverse field ($B_{\rm TF}$) is useful to evaluate the yield of the diamagnetic Mu state (Mu$^+$ or Mu$^-$) by the initial asymmetry distribution in the $\mu$SR time spectra:
\begin{equation}
A(t)\simeq A_{\rm d}e^{-\lambda_\perp t}\cos\omega_\mu t+A_{\rm p}\cos\omega_{\rm p}t,
\end{equation}
where $A_{\rm d}$ is the partial asymmetry of diamagnetic component, $\lambda_\perp$ is the transverse relaxation rate, $\omega_\mu=\gamma_\mu B_{\rm TF}$ with $\gamma_\mu=2\pi\times135.53$ MHz/T being the muon gyromagnetic ratio, $A_{\rm p}$ ($=A_0-A_{\rm d}$, where $A_0$ is the total asymmetry corresponding to 100\% $\mu^+$ polarization) is that of paramagnetic (Mu$^0$) component, and $\omega_{\rm p}=\frac{1}{2}(\gamma_e-\gamma_\mu) B_{\rm TF}$ with $\gamma_e=2\pi\times28.024$ GHz/T being the electron gyromagnetic ratio (which is valid when $\omega_{\rm p}$ is much smaller than the $\mu^+$-$e^-$ hyperfine frequency). In the present measurement ($B_{\rm TF}=2.0$ mT), $\omega_{\rm p}$ ($\simeq2\pi\times28$ MHz) far exceeds the time resolution determined by the muon pulse width ($\delta\approx$80 ns at J-PARC, the corresponding Nyquist frequency being $1/2\delta\approx6.3$ MHz), and the second term is averaged out to yield
\begin{equation}
A(t)\simeq A_{\rm d}e^{-\lambda_\perp t}\cos\omega_\mu t.\label{Atf}
\end{equation}

 The temperature dependence of $A_{\rm d}$ derived from curve fit using Eq.~(\ref{Atf}) is shown in Fig.~\ref{phase}(c). Interestingly, $A_{\rm d}$ in the hexagonal phase shows a pronounced temperature dependence, decreasing significantly from $\sim$0.15 to $\sim$0.05 with the phase transition from the Hex-LT phase to the Hex-HT phase.  The reduction of $A_{\rm d}$ ($<A_0\simeq0.23$) in the hexagonal phase suggests that a part of implanted $\mu^+$s takes a paramagnetic relaxed-excited state, indicating that the band structure with respect to the impurity level associated with Mu is considerably different between the tetragonal and hexagonal phases. This is consistent with the significant difference in the band gap energy between hexagonal and tetragonal/cubic phases inferred from the photo-luminescence energy shown in Fig.~\ref{phase}(b). In the following, we will focus mainly on the diamagnetic Mu and consider the effect of reorientation motion of the FA molecule on the $\mu$SR spectra. 

\begin{figure}[t]
        \begin{center}
                \includegraphics[width=0.95\linewidth,clip]{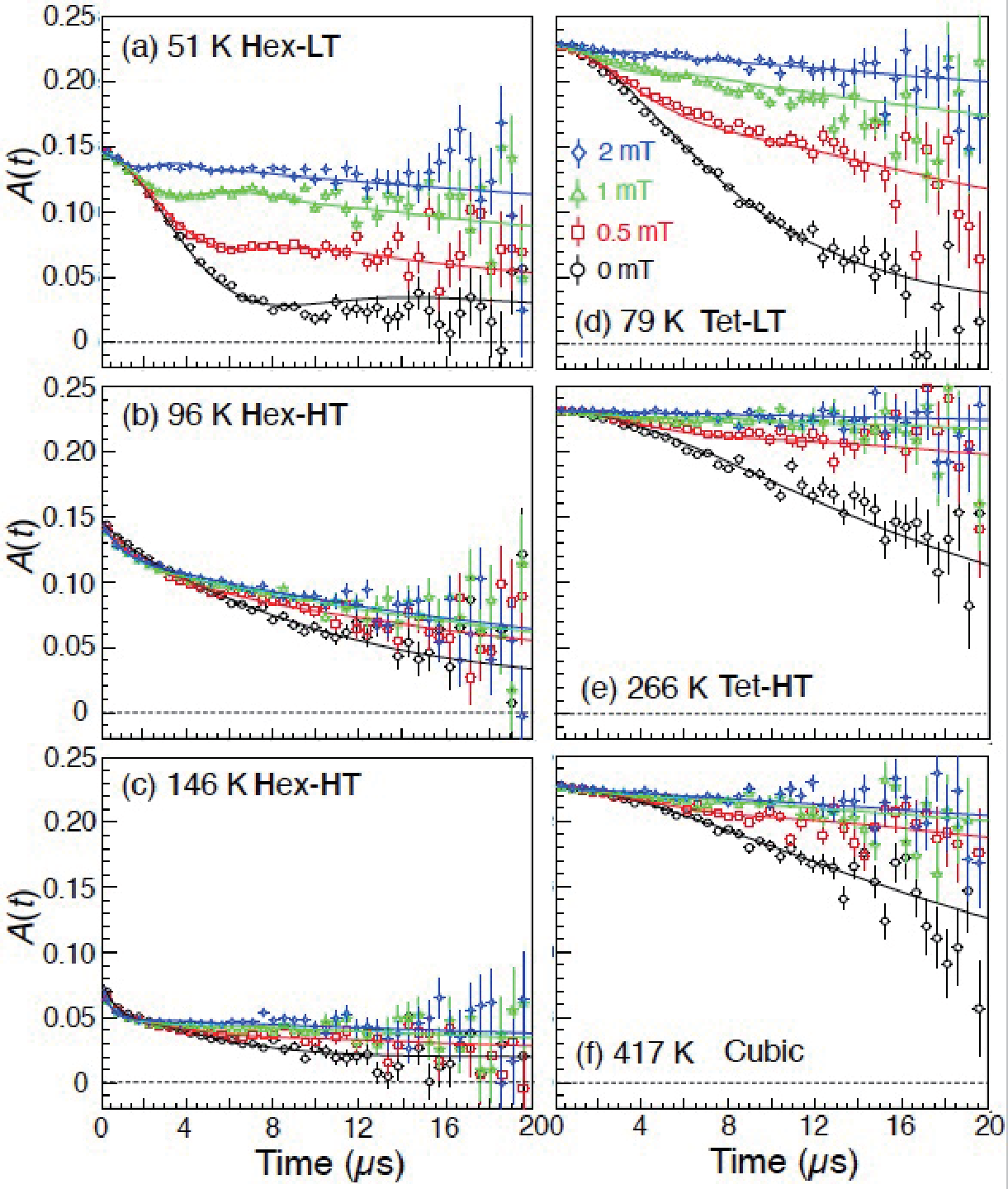}
                \caption{Typical examples of ZF/LF-$\mu$SR time spectra ($\mu$-$e$ decay asymmetry) observed in FAPbI$_3$ at various temperatures. (a)--(c) and (f) are those before quench, and (d), (e) are after quench; see Fig.~\ref{phase} for the corresponding phases.}
\label{tspec}
\end{center}
\end{figure}

Figure \ref{tspec} shows typical ZF/LF-$\mu$SR spectra in the respective phase. These were analyzed by curve fits using the following equation,
\begin{equation}
A(t)\simeq A'_{\rm d}e^{-\lambda_{\rm d}t}G_z^{\rm KT}(t;\Delta,\nu,B_{\rm LF})+A_{\rm p}e^{-\lambda_{\rm p}t}+A_{\rm bg},\label{Alf}
\end{equation}
where $A'_{\rm d}\simeq A_{\rm d}-A_{\rm bg}$, $G_z^\mathrm{KT}(t;\Delta,\nu,B_\mathrm{LF})$ is the dynamical Kubo-Toyabe (KT) function~\cite{Hayano:79}, $\Delta$ is the linewidth determined by random local fields from nuclear magnetic moments, $B_\mathrm{LF}$ is the magnitude of LF,  $\nu$ is the fluctuation rate of $\Delta$, $\lambda_{\rm d}$ and $\lambda_{\rm p}$ are the rate of relaxation induced by fluctuation of hyperfine fields exerted from $\mu^+$-induced charge carriers, and $A_{\rm bg}$ is the background contributed from $\mu^+$s stopped in the sample holder and other parts near the sample in the cryostat.

The KT function is expressed analytically in the case of static ($\nu=0$) and ZF ($B_\mathrm{LF}=0$) conditions,
\begin{equation}
G_z^\mathrm{KT}(t;\Delta,0,0)=\frac{1}{3}+\frac{2}{3}\left(1-\Delta^2t^2\right)e^{-\frac{1}{2}\Delta^2t^2}.\label{KT}
\end{equation}
Provided that all atoms in the crystal are fixed to their lattice points, the magnitude of $\Delta$ for a given Mu site (which we define as $\Delta_0$) is evaluated by calculating a sum of second moments contributed from the $m$-th kind of nuclear magnetic moments ($m = 1,2,3$ and 4 for $^{1}$H, $^{14}$N, $^{127}$I, and $^{207}$Pb with natural abundance $f_m\simeq1$ for $m=1$--3 and $f_4= 0.226$),
\begin{align}
\Delta_0^2&\simeq\gamma_\mu^2\sum_j\langle B_j^2\rangle=\gamma_\mu^2\sum_{m}f_m\sum_{j}\sum_{\alpha=x,y}\sum_{\beta=x,y,z}\gamma_m^2({\bm \hat{D}}_{j}{\bm I}_m)^2 \label{dlts}\\
{\bm \hat{D}}_{j}&=D^{\alpha\beta}_{j}=(3\alpha_{j}\beta_{j}-\delta_{\alpha\beta}r_{j}^2)/r_{j}^5,\quad(\alpha, \beta=x,y,z)\nonumber
\end{align}
where ${\bm r}_{j}=(x_{j},y_{j},z_{j})$  the position vector of the $j$-th nucleus (with Mu at the origin), ${\bm \mu}_m=\gamma_m{\bm I}_m$ the nuclear magnetic moment with $\gamma_m$ being their gyromagnetic ratio. Because $^{14}$N and $^{127}$I nuclei have spin $I_m\ge1$, the corresponding ${\bm \mu}_m$ is subject to electric quadrupolar interaction with the electric field gradient generated by the point charge of the diamagnetic Mu. This leads to the reduction of effective ${\bm \mu}_m$ to the value parallel with ${\bm r}_{j}$ (by a factor $\sqrt{2/3}$ in the classical limit)~\cite{Hayano:79}. From Eq.~(\ref{dlts}), it is immediately understood that the linewidth can be divided into the contribution from the FA molecule, $\Delta_{\rm FA}$, and that of the PbI$_3$ lattice, $\Delta_{\rm PbI}$, as follows:
\begin{equation}
\Delta_0^2 =\Delta_{\rm FA}^2 + \Delta_{\rm PbI}^2.
\end{equation}
As we will show below, the actual value of $\Delta$ (and $\nu$) varies significantly with the reorientation motion of the FA molecule nearby Mu, thus yielding direct information about FA molecular motion.

\begin{figure}[t]
        \begin{center}
                \includegraphics[width=\linewidth,clip]{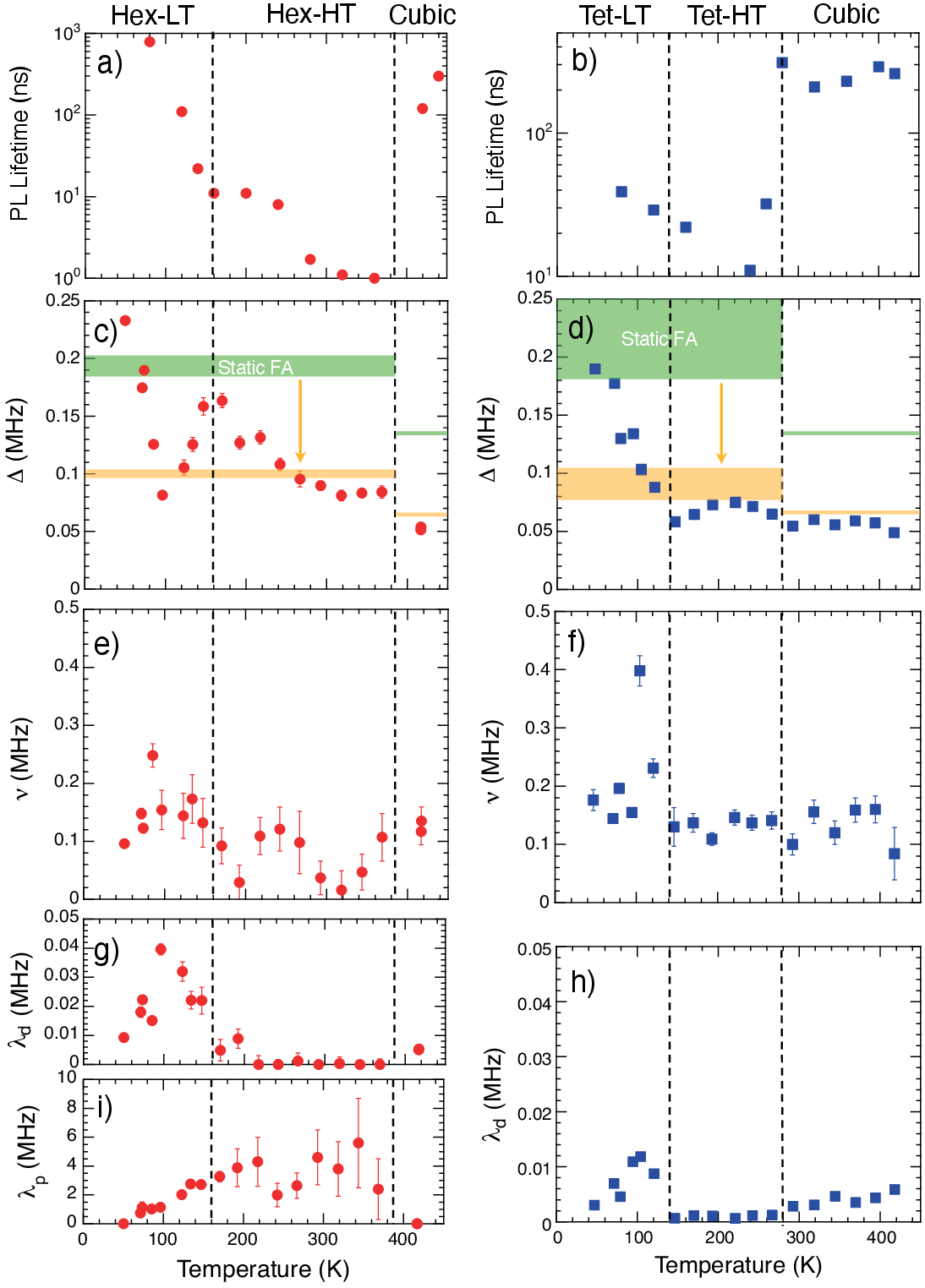}
                \caption{(a), (b) PL lifetime (semi-log plot, quoted from Ref.~[\onlinecite{ChenT:17}]), (c), (d) the Kubo-Toyabe linewidth $\Delta$, (e), (f) fluctuation rate $\nu$, and (g)--(i) exponential relaxation rate versus temperature, where left/right columns represent data on samples before and after quench.  
The horizontal bands hatched in green in (c) and (d) show the expected values when the FA molecule is stationary ($\Delta=\Delta_0$), and those in yellow when the FA contribution disappears due to motional averaging ($\Delta\simeq\Delta_{\rm PbI}$).}
\label{Params}
\end{center}
\end{figure}

We also conducted $\mu$SR measurements under a longitudinal field ($B_{\rm LF}$) up to 2.0~mT at each temperature point. The parameters in Eq.~(\ref{Alf}) were then deduced reliably by global curve fits of the spectra at various $B_{\rm LF}$. $A_{\rm bg}$ was fixed to the value determined by the data at Hex-LT/Tet-LT phases where the depolarization for Mu in the sample was presumed to be complete at later times ($\Delta t\gg1$): $A_{\rm bg}=0.0185$ before quench and $=0.0159$ after quench.  As shown by the solid lines in Fig.~\ref{tspec}, all spectra were coherently reproduced  by the curve fits in the respective phases.  The temperature dependence of the parameters in Eq.~(\ref{Alf}) obtained from this analysis is shown in Fig.~\ref{Params} together with the PL lifetime ($\tau_1$) quoted from Ref.~[\onlinecite{ChenT:17}].

First, we focus on the temperature dependence of the linewidth $\Delta$ in the sample before quench. As shown in Fig.~\ref{Params}(c), $\Delta$ shows a maximum at the lowest temperature of the Hex-LT phase, decreases rapidly with increasing temperature up to $\sim$100 K, then begins to increase to reach a maximum around 160 K at the boundary with the Hex-HT phase, and decreases again with increasing temperature up to $\sim$300 K. This is qualitatively the same as the $\mu$SR results in MAPbI$_3$, suggesting that $\Delta_{\rm FA}$ (which is mainly carried by the protons) is reduced due to motional averaging by their reorientation motion. The origin of enhanced $\Delta$ around the phase boundary is also assumed to be the same as in the case of MA. Namely, the LT-HT phase transition is induced by the thermally agitated reorientation motion of the FA molecules, and the interaction between FA and the PbI$_3$ lattice resonantly increases near the phase boundary, causing a slowdown. (In the case of MA, this scenario is also supported by the fact that the structural phase transition temperature shifts significantly toward the higher temperature side due to the deutration of the MA molecules.) The above interpretation is quantitatively supported by the values of $\Delta_0$ for the Mu sites deduced from DFT calculations for each phase evaluated using Eq.~(\ref{dlts}): as seen in Fig.~\ref{model}(c), $\Delta$ at the lowest temperature is close to $\Delta_0$ in the presence of stationary FA molecules ($\sim$0.2 MHz, indicated by a green band, where the width in the Hex-LT/HT phases represents the ambiguity due to the muon site), whereas in the higher temperature Hex-Cubic phases it approaches $\sim$0.1 MHz without FA molecules ($\simeq\Delta_{\rm PbI}$, yellow band). 

The parameter $\nu$, which gives the fluctuation frequency of $\Delta$ in the KT relaxation function over a time range of $10^{-8} <\nu^{-1}<10^{-4}$ s (where $\Delta$ serves as a constant), does not show a pronounced temperature dependence, as seen in Fig.~\ref{Params}(e). This result is in apparent contrast to the increase in $\nu$ observed near the phase boundary at low temperatures seen in the MA case. However, as is clear from Figs.~\ref{phase} and \ref{tspec}, the partial asymmetry of the diamagnetic Mu ($A_{\rm d}$) decreases rapidly toward the phase boundary in the Hex-LT/HT phase that coincides with the reduction of $\Delta$, which makes it extremely difficult to observe the change in the KT relaxation function due to the $\nu$ change in this temperature region. Therefore, the lack of sufficient sensitivity to $\nu$ in the Hex-LT/HT phase is considered to be responsible for the apparent difference with MA. A similar situation would be true for $\lambda_{\rm d}$, which represents the effect of paramagnetic spin fluctuations on $\Delta$: the monotonous increase of $\lambda_{\rm p}$, which reflects the paramagnetic spin fluctuations themselves, is also consistent this interpretation.

Next, let us look at the results for the sample after quench into the tetragonal structures. In contrast to the results before quench, $\Delta$ decreases monotonically with increasing temperature in the Tet-LT phase and approaches the value in the Tet-HT phase; $\Delta$ exhibits almost no temperature change in the Tet-HT and cubic phases, and its value is almost consistent with the value expected in the absence of FA molecules ($=\Delta_{\rm PbI}$, yellow band). In the Tet phase, the evaluation of $\Delta_0$ is highly arbitrary because of the large number of inequivalent FA molecular configurations reflecting the relative ease of reorientation for FA: $\Delta_0$ varies from 0.18 to 0.29 MHz depending on whether different FA configurations are taken into account by the virtual crystal approximation or not [as indicated by a green band in Fig.~\ref{Params}(d)]. In any case, $\Delta$ at the lowest temperature is consistent with $\Delta_0$ where the FA molecules appear to be nearly stationary.

The temperature dependence of $\nu$ is similar to that of MA, showing a peak at $\sim$100 K where $\Delta$ exhibits a rapid decrease. This is probably because the yield of diamagnetic Mu in the quenched sample is close to 100\% regardless of temperature, as seen in Figs.~\ref{phase} and \ref{tspec}, which is common to the MA case. Conversely, the temperature dependence of $\lambda_{\rm d}$ indicates that its sensitivity to paramagnetic spin fluctuations decreases in proportion to $\Delta$ (which is similar to the case in the sample before quench).

\section{Discussion}
In a previous study using $^1$H-NMR on a sample that seems to correspond to the post-quench phase, the relaxation time of FA molecular reorientation derived from the longitudinal relaxation rate ($1/T_1$) was reported to decrease rapidly from $\tau_{\rm r}\sim$10$^{-8}$ s to $4\times10 ^{-11}$s (at 140 K) in the temperature region corresponding to the Tet-LT phase, followed by a moderate decrease to $1\times10^{-11}$s in the Tet-HT and cubic phases \cite{Fabini:16,Fabini:17}.  Assuming that this causes fluctuations of $\Delta_{\rm FA}$ so that $\nu\simeq1/\tau_{\rm r}$, we would expect $\nu\gg\Delta_{\rm FA}$ and $\Delta_{\rm FA}$ is effectively reduced to zero due to motional averaging. Therefore, the sharp decrease of $\Delta$ in the Tet-LT phase corresponds to the reduction of $\tau_{\rm r}$, and is consistent with our interpretation that the value of $\Delta$ above 140 K corresponds to $\Delta_{\rm PbI}$. 
 We also speculate that the slow fluctuations of $\Delta$ observed in $\mu$SR ($1/\nu\simeq10^{-5}$ s), except for the peak observed at temperatures near 100 K for the sample after quench, are of a different origin than the reorientation of the FA molecules.

\begin{figure}[t]
        \begin{center}
                \includegraphics[width=\linewidth,clip]{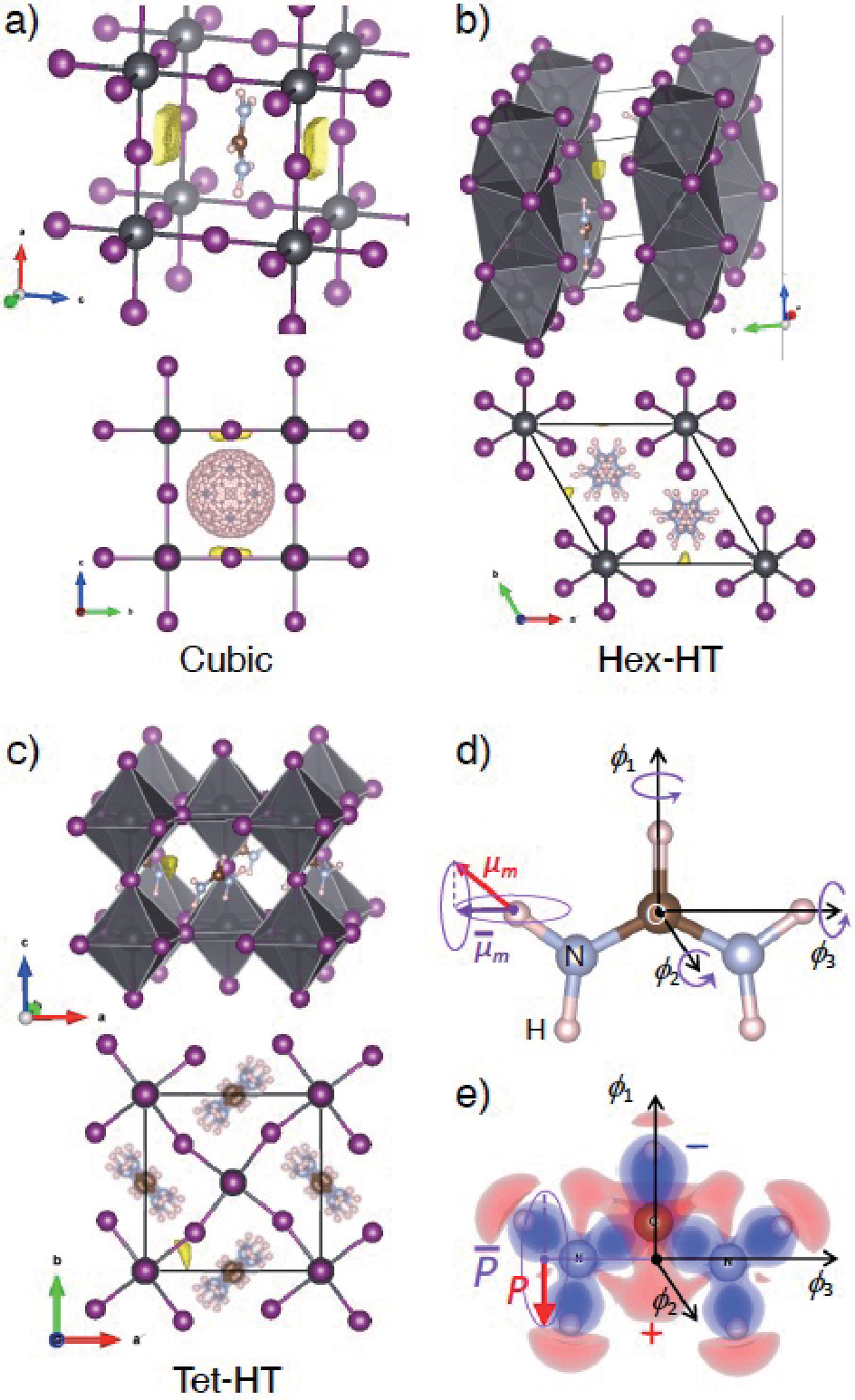}
                \caption{Crystal structure of FAPbI$_3$ for (a) cubic, (b) Hex-HT, and (c) Tet-HT phases, where the yellow-hatched areas show the Mu sites estimated by DFT calculations. (d) Schematic illustrations of FA molecule with three symmetry axes ($\phi_i$, $i=1,2,3$) of jumping rotation. The contribution of nuclear magnetic moments ($\mu_m$) to $\Delta$ is effectively reduced by the reorientation around the $\phi_3$ axis to $\overline{\mu}_m$ by motional averaging (a projection to the $\phi_3$ axis) and then to zero by further averaging around $\phi_{1,2}$ axes. (e) A mechanism of reducing electric dipoles (${\bm P}\rightarrow\overline{\bm P}\approx0$) similar to that for $\mu_m$ is illustrated, where the hatched areas indicate local charge asymmetry obtained by DFT calculation \cite{ChenT:17}.}
\label{model}
\end{center}
\end{figure}

The previous DFT simulations of FA molecular rotation showed that the cubic phase has the lowest potential barrier to rotation around all three symmetry axes shown in Fig.~\ref{model}(d), while the rotation is restricted due to the higher potential barrier around the $\phi_{1,2}$ axes (0.2--0.6 eV) in the Tet-LT/HT phases~\cite{Fabini:17}.
As is illustrated in Fig.~\ref{model}(d), the fast reorientation around the $\phi_3$ axis reduces the contribution of $\mu_m$ for $^1$H and $^{14}$N nuclei, reducing $\mu_m$ to $\overline{\mu}_m$ (corresponding the projection of $\mu_m$ to the $\phi_3$ axis). Specifically,
the nuclear dipolar fields in Eq.~(\ref{dlts}) are expressed as
\begin{equation}
B_{j}=\frac{\mu_m}{r_{j}^3}[(3\cos^2\theta_{\bm r}-1)\cos\theta_{m}+
3\sin\theta_{\bm r}\cos\theta_{\bm r}\sin\theta_{m}\cos\psi_{m}],\label{hdip}
\end{equation}
where  $\theta_{\bm r}$ is the polar angle of ${\bm r}_{j}$, $\theta_{m}$ and $\psi_m$ are the polar and azimuth angle of ${\bm \mu}_m$. Assuming that the $\hat{z}$ axis for the polar coordinates is parallel with $\phi_3$, the term proportional to $\sin\theta_{m}$  in Eq.~(\ref{hdip}) is averaged out by the jumping rotation with remaining contribution $\overline{\mu}_m=\mu_m\cos\theta_{m}$ when $\tau_3^{-1}\gg\Delta_{\rm FA}$. While the extent of decrease in $\Delta$ due to this depends on the relationship between the $\phi_3$ axis and Mu site, it is estimated to decrease by a factor of $1/\sqrt{2}$ in the random average \cite{Hayano:79}. The FA contribution is eventually eliminated by further averaging around $\phi_{1,2}$ axis ($\perp \phi_3$) when $\tau_{1,2}^{-1}\gg\Delta_{\rm FA}$. Thus, provided that the difference in activation temperatures of reorientation motion around the $\phi_{3}$ and $\phi_{1,2}$ axes is large enough, $\Delta$ is expected to exhibit decrease in two steps with increasing temperature. However, given that the contribution from the FA molecules to $\Delta$ almost disappears at around 140 K [see Fig.~\ref{Params}(d)], it can be inferred that the potential barrier for the rotation is relatively small for all the three axes in Tet-LT/HT phase. Since the DFT calculations do not take into account the relaxation of PbI$_3$ lattice \cite{Fabini:17}, this suggests that the lattice relaxation occurs to the extent that it does not interfere with the rotation of the FA molecules. 

In contrast, the observed temperature dependence of $\Delta$ in the Hex-LT/HT phase [Fig.~\ref{Params}(c)] is qualitatively similar to that in the case of MAPbI$_3$ in that it occurs in almost two steps with a maximum at the boundary of the orthorhombic and tetragonal phases: $\Delta$ decreases with increasing temperature due to reorientation associated with rotation around the $C_3$ symmetry axis (corresponding to the $\phi_3$ axis for FA molecules) in the tetragonal phase, and then reorientation around the $C_4$ symmetry axis (corresponding to the $\phi_{1,2}$ axes) develops rapidly near the structural phase transition temperature to the tetragonal phase ($\sim$162 K), and $\Delta$ reaches a maximum once and then decreases gradually. Therefore, it can be inferred that almost the same scenario holds in the Hex-LT/HT phase for FA. This suggests that, in contrast to the Tet-LT/HT phase, the reorientation motion of FA molecules is relatively strongly restricted by the PbI$_3$ lattice.

Let us look at the relationship between $\tau_{\rm PL}$ (corresponding to $\tau_{\rm c}$), and $\Delta_{\rm FA}$ [$=(\Delta-\Delta_{\rm PbI})^{1/2}$], which is now interpreted as proportional to $\tau_{\rm r}$ for the FA molecules. First, it is noteworthy that the overall temperature dependence of $\Delta$, including the hump around 160 K in the sample before quench, is very similar to that observed for $\tau_{\rm PL}$~\cite{ChenT:17}. As shown in Fig.~\ref{Params}(a) (semi-log plot), the longer $\tau_{\rm PL}$ ($=\tau_1$ in Ref.~[\onlinecite{ChenT:17}]) decreases rapidly above $\sim$80 K, stops at around 160--220 K, and then decreases rapidly above 220 K.  [A similar trend is observed for the shorter lifetime ($\tau_2$, not shown)]. This is qualitatively in line with the case of MAPbI$_3$ \cite{Koda:22}. On the other hand, in the sample after quench, $\tau_{\rm PL}$ is already more than one order of magnitude shorter than that before quench at around 80 K [the lowest temperature of the data, see Fig.~\ref{Params}(b)], consistent with a more rapid decrease in $\Delta$ in the Tet-LT phase than in the Hex-LT phase. Such a correlation suggests an intrinsic relationship between $\tau_{\rm c}$ and FA molecular motion, as in the case of MA, and supporting the idea that the mechanism causing the change in $\Delta$ is also acting on the local dielectric permittivity $\varepsilon_{\rm loc}(\omega)=\varepsilon'_{\rm loc}(\omega)+i\varepsilon''_{\rm loc}(\omega)$.  As illustrated in Fig.~\ref{model}(e), the FA molecule is accompanied by an electric dipole moment  (${\bm P}=0.35$ Debye) along the $\phi_1$ axis due to local charge imbalance \cite{ChenT:17}, and $\varepsilon''_{\rm loc}(\omega\approx\tau_c^{-1})$ is expected to decrease for $\tau_{\rm r}\ll\tau_{\rm c}$ by motional averaging as in $\Delta$\:~\cite{Koda:22}. However, since ${\bm P}$ has only one component parallel to the $\phi_1$ axis, it differs from the $\Delta$ case in that it becomes zero by fast jump rotation around either the $\phi_2$ or $\phi_3$ axis. Thus, the anomaly shown by $\Delta$ and $\tau_{\rm PL}$ around 160 K suggests that the activation of the rotational mode around the $\phi_{1,2}$ axis temporarily inhibits the reorientation around the $\phi_3$ axis in the process of inducing the structural phase transition.

The implication that the more reorientation of the cation molecule is activated, the less the photo-excited carriers are able to form the large polarons required for longer $\tau_{\rm c}$ suggests that, since such activation is a random thermal excitation by phonons, the effect is rather that of disrupting the coherent motion of the cation molecules necessary for large-polaron formation. NMR and neutron scattering suggest that  $\tau_{\rm r}\sim10^{-7}$--$10^{-8}$ s at temperatures below 50 K where $\tau_{\rm c}$ is long. In other words, the cation molecules can rotate even at such low temperatures and are not strongly scattered by phonons, and thus can reorient coherently with respect to the carriers. Based on these considerations, it may be useful to search for combinations of cation molecules and PbX$_3$ lattices that can suppress the fast reorientation of molecules as a guideline for material design to achieve long $\tau_{\rm c}$.

Finally, we briefly discuss the origin of the PL lifetime recovery in the cubic phase. It is clear from Figs.~\ref{Params}(a) and (b) that the increase in $\tau_{\rm PL}$ in this phase is completely uncorrelated with $\Delta$, both before and after quench. Therefore, the cause of the increase in $\tau_{\rm PL}$ is independent of the degrees of freedom of the FA molecule. This is consistent with the high photoelectric conversion efficiency of CsPbI$_3$ in the same phase, where MA/FA cations are replaced with Cs ions \cite{Wang:23}. Interestingly, the real part of the bulk permittivity $\varepsilon'(\omega)$ shows a clear increase from low temperatures toward the boundary between the Tet-HT and cubic phases at $\sim$280 K ($\varepsilon'=44\rightarrow46$) and further increases with increasing temperature \cite{Cordero:19}. A similar trend is reported for FAPbCl$_3$ \cite{Govinda:18} and CsPbBr$_3$ \cite{Svirskas:20}. These facts may imply that the recovery of $\tau_{\rm PL}$ is related to $\varepsilon'(\omega)$ derived from the displacement of the center of gravity of the cation.

\section*{ACKOWLEDGMENTS}
We would like to thank the MLF staff for their technical support. This work was supported by the Elements Strategy Initiative to Form Core Research Centers, from the Ministry of Education, Culture, Sports, Science, and Technology of Japan (MEXT) under Grant No.~JPMXP0112101001, and partially by the MEXT Program: Data Creation and Utilization Type Material Research and Development Project under Grant No.~JPMXP1122683430. M.H. also acknowledges the support of JSPS KAKENHI Grant No.~19K15033 from MEXT.  The $\mu$SR experiments were conducted at the Materials and Life Science Experimental Facility (MLF), J-PARC under the support of Inter-University-Research Programs (Proposals No. 2018MI21) by Institute of Materials Structure Science, KEK. S.-H. L. and J. J. C. acknowledge support from the U.S. Department of Energy, Office of Science, Office of Basic Energy Sciences under Award Number DE-SC0016144.
\vspace{1cm}
\input{fapi.bbl}
\end{document}

%% file: fapi.bbl
%